\documentstyle[referee]{mn}
\newcommand{\reference}{\bibitem}
\def\beq{\begin{equation}}
\def\eeq{\end{equation}}
\def\bey{\begin{eqnarray}}
\def\eey{\end{eqnarray}}
\def\beqarray{\begin{eqnarray}}
\def\eeqarray{\end{eqnarray}}

\def\mpc{\,{\rm {Mpc}}}
\def\Mpc{\,{\rm {Mpc}}}
\def\kpc{\,{\rm {kpc}}}
\def\kpch{\,{h^{-1}{\rm kpc}}}
\def\mpch{\,h^{-1}{\rm {Mpc}}}
\def\kms{\,{\rm {km\, s^{-1}}}}
\def\msun{M_\odot}

\def\v200{V_{200}}

\def\md{m_d}
\def\Onow{\Omega_0}
\def\Lnow{\Lambda_0}
\def\Rd{R_d}
\def\my{{M_\odot \rm yr^{-1}}}

\input epsf
\title[]	
{The Structure and Clustering of Lyman Break Galaxies}

\author[]	
{H. J. Mo, Shude Mao, Simon D. M. White
\thanks {E-mail: (hom, smao, swhite)@mpa-garching.mpg.de} \\
      Max-Planck-Institut f\"ur Astrophysik
      Karl-Schwarzschild-Strasse 1, 85748 Garching, Germany}
\date{Accepted ........
      Received .......;
      in original form .......}
\pubyear{1998}

\begin{document}
\maketitle
\begin{abstract}
The number density and clustering properties of Lyman-break 
galaxies (LBGs) are consistent with them being the 
central galaxies of the most massive dark halos present at $z\sim 3$. 
This conclusion holds in all currently popular 
hierarchical models for structure formation, and is almost independent
of the global cosmological parameters. We examine whether the sizes, 
luminosities, kinematics and star-formation rates of LBGs are also 
consistent with this identification. Simple formation models tuned to 
give good fits to low redshift galaxies can predict the distribution 
of these quantities in the LBG population. The LBGs should be small 
(with typical half-light radii of $0.6-2\kpch$), should inhabit
haloes of moderately high circular velocity ($180-290 \kms$) but
have low stellar velocity dispersions ($70-120 \kms$) and should
have substantial star formation rates ($15-100$M$_\odot$yr$^{-1}$).
The numbers here refer to the predicted median values
in the LBG sample of Adelberger et al. (1998); the first assumes an
$\Omega_0=1$ universe and the second a flat
universe with $\Omega_0=0.3$. For either cosmology these predictions 
are consistent with the current (rather limited) observational data.
Following the work of Kennicutt (1998) we assume stars to form more
rapidly in gas of higher surface density. This predicts that LBG
samples should preferentially contain objects with low angular momentum, 
and so small size, for their mass. In contrast, samples of damped Ly$\alpha$ 
systems (DLSs), should be biased towards objects with large angular
momentum. Bright LBGs and DLSs may therefore form distinct populations,
with very different sizes and star formation rates, LBGs being 
smaller and more metal-rich than DLSs of similar mass and redshift.
\end{abstract}

\section {Introduction}

The Lyman-break technique is remarkably effective in finding
galaxies at $z\sim 3$ bright enough 
for spectroscopy at the Keck telescope. Redshifts are now 
available for almost 700 systems to an optical R-band magnitude 
of about 25.5 (Steidel, Pettini \& Hamilton 1995; Steidel et 
al 1996; Steidel et al. 1998a,b; Adelberger et al. 1998). This
sample is flux-limited in the rest frame UV, implying a lower 
limit on the star-formation rates of the observed galaxies.
The comoving density of these Lyman-break galaxies (LBGs) is comparable  
to that of present-day bright galaxies. Based on this and on the 
equivalent widths of the saturated absorption lines, Steidel et al. 
argued that these LBGs are probably the progenitors of the spheroids
of luminous galaxies. This conclusion is tentative,
however, since the observed equivalent widths appear strongly 
affected by outflows and so may not be associated with deep 
potential wells.

Mo \& Fukugita (1996) noted that the abundance and linewidth of LBGs
may provide important constraints on theories of structure formation. 
Assuming the LBGs to be associated with the most massive haloes 
present at $z\sim 3$, they showed that in many (but not all) popular
cosmogonies the host haloes would have circular velocities
$\ga 200\kms$, comparable to the velocity dispersions inferred from
the observed line widths. As we will see below, this agreement is probably
a fluke -- the observed line widths plausibly overestimate the
stellar velocity dispersion, and the latter should, in any case, 
be substantially smaller than halo circular velocity. The power of
this association, however, is that it
makes specific predictions for the clustering properties of
the LBG population. Mo \& Fukugita pointed out that massive
haloes should be much more strongly clustered than the underlying mass at 
$z\sim 3$. Thus the LBGs and their descendents should show stronger 
spatial correlations than less luminous galaxies. This is a generic 
prediction of hierarchical structure formation and depends little
on the details of how LBGs form. The clustering of LBGs thus provides
a test both of the hierarchical model and of the identification of the
LBGs with the most massive haloes. 

Recent papers presenting data for large LBG samples have used
this simple hypothesis to show that the relatively strong clustering
they measure is consistent with the predictions of hierarchical
clustering models both for high and for low density cosmologies
(Steidel et al. 1998a,b; Giavalisco et al. 1998; Adelberger et al. 1998). 
Indeed, the observed clustering strength was predicted in advance with
remarkable accuracy by the semi-analytic models of Baugh et al.
(1998). These models follow galaxy formation in a hierarchical 
cosmology in some detail; they suggest that for the current 
spectroscopic LBG
samples it may be a good approximation to assume that each dark halo 
contains a  single LBG with a star-formation rate depending primarily 
on halo mass. Since publication of the clustering measurements,
a number of theoretical studies have used analytic methods,
pure N-body simulations, N-body simulations combined with
semi-analytic galaxy formation models, and full N-body +
hydrodynamics simulations to argue that the observed clustering 
at $z\sim 3$ is easily reproduced in most currently popular CDM
cosmologies (Bagla 1998a,b; Coles et al. 1998; Governato et al. 1998;
Jing 1998; Jing \& Suto 1998; Katz, Hernquist \& Weinberg 1998; 
Moscardini et al. 1998; Peacock et al. 1998; Wechsler et al. 1998). 
This should not be
surprising. Even the first simulation of galaxy 
clustering in a CDM universe showed that correlations of
bright galaxies should evolve very slowly in comoving
coordinates even though evolution of the mass correlations is strong
(cf. Fig.17 in Davis et al. 1985). 

The assumption that LBGs are the central galaxies of massive haloes 
provides a framework for predicting a variety of other
observables. Comparing these with the data then gives further tests of
the underlying galaxy formation paradigm. In the present paper we 
use the Press-Schechter (1974) model to predict the abundance of dark 
haloes as a function of mass and redshift; we adopt the analytic 
fitting formulae of Navarro, Frenk \& White (1997) to specify the 
internal density structure of haloes; we follow Mo, Mao \& White 
(1998) in assuming that central galaxies form when collapse of the 
protogalactic gas is arrested either by its spin, or by fragmentation 
as it becomes self-gravitating; and we use the empirical results of 
Kennicutt (1998) to determine star formation rates. Section 2 below 
presents details of these models and explores the consequences of 
assuming that LBGs are the central galaxies of the {\it most 
massive} haloes. This allows us to confirm previous work in the 
context of our own models and to clarify how our later results 
should scale as cosmological parameters vary. In \S 3 we 
predict sizes, kinematics, star formation rates and halo
masses for LBGs based on the hypothesis that the observed 
samples correspond to the {\it most rapidly star-forming} central 
galaxies at $z\sim 3$. We also compare the predicted properties of this
population to those of damped 
Ly$\alpha$ absorbers at the same redshift. In \S 4, we discuss 
our results further and summarise our conclusions.

\section {Modelling Lyman Break Galaxies}

We model the assembly of galaxies in the context of the standard
hierarchical picture (e.g. White \& Rees 1978; Blumenthal et al. 1984; 
White \& Frenk 1991; Kauffmann, White \& Guiderdoni 1993; Cole et al. 
1994). Structure growth in these models is specified by the
parameters of the background cosmology and by the power spectrum
of initial density fluctuations. The relevant cosmological parameters 
are the Hubble constant $H_0=100h\kms\mpc ^{-1}$, the total matter 
density $\Omega_0$, the mean density of baryons $\Omega_{\rm B}$, and the 
cosmological constant $\Lambda$. The last three are all expressed in 
units of the critical density for closure. We will use models within 
the cold dark matter (CDM) family. The power spectrum $P(k)$ is then
specified by an amplitude, conventionally quoted as 
$\sigma_8$, the {\it rms} linear overdensity at $z=0$ in spheres of 
radius $8h^{-1}$Mpc, and by a shape parameter $\Gamma$. We do not
consider the possibility of a tilt and we neglect the weak dependence
of $P(k)$ on baryon density.

For a given cosmogony, we estimate the mass function of dark matter 
haloes (their abundance as a function of mass)
from the Press-Schechter formalism (Press \& Schechter 1974, PS):
\beq\label{PS}
{dN_{\rm h}\over dM_{\rm h}}(M_{\rm h},\, z)~dM_{\rm h}
=-\sqrt{2\over\pi} {{\overline \rho}_0 \over M_{\rm h}}
{\delta_c(z)\over \Delta (M_{\rm h})}{d\ln \Delta (M_{\rm h})\over d\ln M_{\rm h}}
\exp\left[-{\delta_c^2(z)\over 2\Delta^2 (M_{\rm h})}\right]
{dM_{\rm h}\over M_{\rm h}},
\eeq
where $\delta_c(z) = \delta_c(0)(1+z)g(0)/g(z)$ is the linear overdensity
corresponding to collapse at redshift $z$, $\delta_c(0)\approx
1.686$, $g(z)$ is the linear growth factor relative to an Einstein-de
Sitter universe,
$\Delta(M_{\rm h})$ is the {\it rms} linear mass fluctuation
at $z=0$ in a sphere which on average contains mass $M_{\rm h}$, and
${\overline \rho}_0$ is the mean mass density of the universe
at $z=0$. The circular velocity and virial radius of a halo are determined
from its mass and redshift according to
\beq\label{VhM}
V_{\rm h}^2={GM_{\rm h} \over r_{\rm h}}, \,\,\,\,\, 
r_{\rm h}=\left[{GM_{\rm h}\over 100H^2(z)}\right]^{1/3},
\eeq
where $H(z)$ is the Hubble constant at redshift $z$.
A more detailed description of this formalism and of 
related issues can be found in the Appendix of Mo, Mao \& White (1998).

\subsection {Lyman Break Galaxies as the Most Massive Haloes at $z\sim 3$}

Many of the properties we predict for the LBG population can be
understood using a very simple model which we now describe.
Suppose that each massive halo at $z\sim 3$
has a central galaxy with a star formation rate (SFR) which 
is a monotonic function of halo mass. Suppose further that the
rest-frame UV luminosities of these galaxies increase monotonically with
their SFR. Suppose finally that only a negligible fraction of haloes host a
second galaxy bright enough to be seen. The observed LBGs then
correspond to the most massive haloes at $z\sim 3$ and
the sample magnitude limit corresponds to a lower limit on halo 
circular velocity. We can estimate this limit by
calculating the abundance of massive haloes from
equations (1) and (2) and equating it to the observed abundance
of LBGs. For the latter we adopt the number given by Adelberger et
al (1998) for LBGs brighter than ${\cal R}=25.5$. This is 
$N_{\rm LBG}\approx 8\times 10^{-3} h^3\Mpc^{-3}$ at 
$z\sim 3$ for an Einstein-de Sitter universe, and is
quite similar to the present abundance of $L_*$ galaxies.
When considering other cosmologies, we estimate the appropriate 
observed number density by dividing this number by the comoving 
volume per unit redshift at $z=3$ and multiplying by the corresponding
value for an Einstein-de Sitter (EdS) universe.

\begin{figure}
\epsfysize=9.5cm
\centerline{\epsfbox{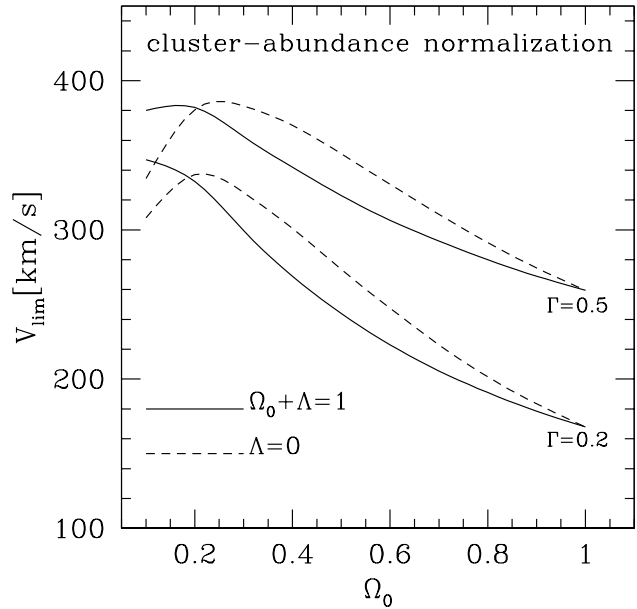}}
\caption{The lower limit on $V_{\rm h}$ 
required to reproduce the observed number density
of LBGs. Results are shown for flat and open CDM models 
with shape parameters of $\Gamma=0.5$ and 0.2. The fluctuation 
amplitudes in all models are normalised to produce the
observed cluster abundance at $z=0$.
}\end{figure}

Figure 1 shows how the minimum halo circular velocity derived in this
way depends on the cosmology assumed. Results are given for flat
and open cosmologies and for CDM power spectra with $\Gamma=0.2$ and 
0.5. In all cases we normalize the power spectra according to the
observed abundance of rich clusters at $z=0$; we take this to
require $\sigma_8=0.6 \Omega_0^{-0.55}$ (White, Efstathiou \& Frenk
1993; Viana \& Liddle 1996). 
For given $\Gamma$, the limiting circular velocity generally 
increases with decreasing $\Omega_0$ because big haloes
then form earlier. This trend reverses
below $\Omega_0\sim 0.2$ in the open sequence; in such models
most of the mass is already part of a few massive haloes by $z=3$.
For a given $\Omega_0$, $V_{\rm lim}$ is higher for larger
$\Gamma$ because this implies higher amplitude fluctuations 
on galactic scale given the fixed amplitude on cluster scale. 
As shown in the figure, models with $\Omega_0\la 0.3$ have 
$V_{\rm lim}\ga 300\kms$, corresponding to a total halo mass 
$M_{\rm h}\sim 1.0\times 10^{12}h^{-1}\msun$. 
In such models, LBGs are indeed associated with massive dark 
haloes. In contrast, for $\Gamma=0.2$ and $\Omega_0\sim 1$ 
we find $V_{\rm lim}\sim 170\kms$, corresponding to $M_{\rm h}\sim 
1.5\times 10^{11} h^{-1}\msun$. For these parameters few
massive haloes form before $z=3$ and one has to 
include smaller haloes in order to match the observed number 
density of LBGs.

\begin{figure}
\epsfysize=9.5cm
\centerline{\epsfbox{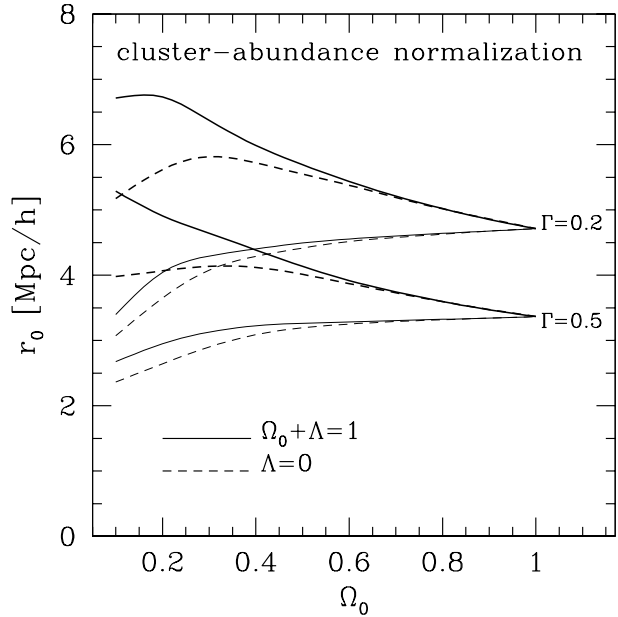}}
\caption{The correlation lengths predicted for
LBGs in the cosmological models of Fig~1. 
Thick lines (the upper pair for each $\Gamma$)
show the physical correlation length in comoving units,
while thin lines (the lower pair for each $\Gamma$)
show the scaled correlation length,
$r_0 D_\star (1, 0)/D_\star (\Omega_0, \Lambda)$, 
where $D_\star(\Omega_0, \Lambda)$ is the angular size distance 
at $z=3$. This is a constant times the angular scale corresponding to
$r_0$ and so any comparison with observations depends very weakly
on the cosmology assumed when analysing the observational data.
}\end{figure}

We can define a characteristic halo circular velocity
at any redshift by requiring the {\it rms} linear density
fluctuation on the corresponding mass scale to be $\delta_c(z)$. 
At $z=3$ the limiting circular velocities 
obtained above are much larger than this characteristic value
for all models except the open sequence at $\Omega_0\la 0.2$.
As a result, the distribution of the LBG haloes is 
strongly biased relative to that of the mass.
We can calculate this bias 
explicitly using the model of Mo \& White (1996), according to which
the halo two-point correlation function (and hence the LBG correlation
function) can be written as
\beq
\xi_{\rm h}(r)=b^2 \xi_{\rm m} (r),
\eeq
where $\xi_{\rm m}$ is the correlation function for the mass.
For haloes with circular velocity $V_{\rm h}$ (corresponding
to mass $M_{\rm h}$) the bias parameter can be written as  
\beq\label{bias}
b(V_{\rm h}, z)=1+{1\over \delta_c(0)}
\left[ {\delta_c^2(z)\over \Delta^2 (M_{\rm h})}-1\right].
\eeq
For the LBG population as a whole, the appropriate bias factor 
is an average of $b(V_{\rm h}, z)$ weighted by abundance as a
function of $V_{\rm h}$ (taken here from equations 1 and 2). 
The mass correlations $\xi_{\rm m}$ can be calculated
analytically as described in Mo, Jing \& B\"orner (1997;
see also Jain, Mo \& White 1995; Peacock \& Dodds 1996).
Equation (3) can then be used to estimate the correlation length
of LBG galaxies using the definition $\xi_{\rm h}(r_0)=1$.

The thick lines in Figure 2 show the values of $r_0$ obtained in
this way for the sequences of cosmologies already
studied in Figure 1. For given $\Gamma$ the correlation length 
increases with decreasing $\Omega_0$ 
except for the open sequence at $\Omega_0\la 0.3$. This behaviour is the
result of two competing effects. 
The mass correlations at $z=3$ are stronger in low $\Omega_0$ 
universes because structures grow more slowly with time. On the other 
hand, the bias factor is lower for low $\Omega_0$ because 
$\delta_c(z)$ is smaller and $\Delta$ is larger (reflecting 
the larger value of $\sigma_8$).

The observational estimate of $r_0$ for LBGs depends on the assumed
cosmology because the angular size distance is needed to convert
angular separations into physical distances. Based on 
a count-in-cells analysis, Adelberger et al. (1998) found 
$r_0=(4\pm 1)\mpch$ under the assumption of an EdS
universe and $r_0=(6\pm 1)\mpch$  for an open universe with 
$\Omega_0=0.2$. Both these values are consistent with our model 
predictions if $\Gamma \sim 0.2$. 

An observational estimate of clustering amplitude
which depends very weakly on the assumed cosmology can be 
obtained by dividing $r_0$ by a typical angular size distance 
in order to convert it to the corresponding angular scale. We show
predictions for this estimate in Figure 2, where it has been multiplied 
by the typical angular size distance in the EdS case in order to
convert to the same units used for $r_0$ itself.
The angular scale corresponding to $r_0$ is predicted to depend 
rather little on $\Omega_0$ and $\Lambda$, but more strongly on
$\Gamma$. Adelberger et al. (1998) obtained a value $(4\pm 1)\mpch$
for this scaled quantity, consistent with the $\Gamma=0.2$
models of Figure 2 for all $\Omega_0$ and with the
$\Gamma=0.5$ models for all but the smallest $\Omega_0$.
The data clearly support the underlying theoretical paradigm, but
it appears that error bars at least a factor of two smaller will be 
needed to get significant constraints on cosmological parameters.

\begin{figure}
\epsfysize=9.5cm
\centerline{\epsfbox{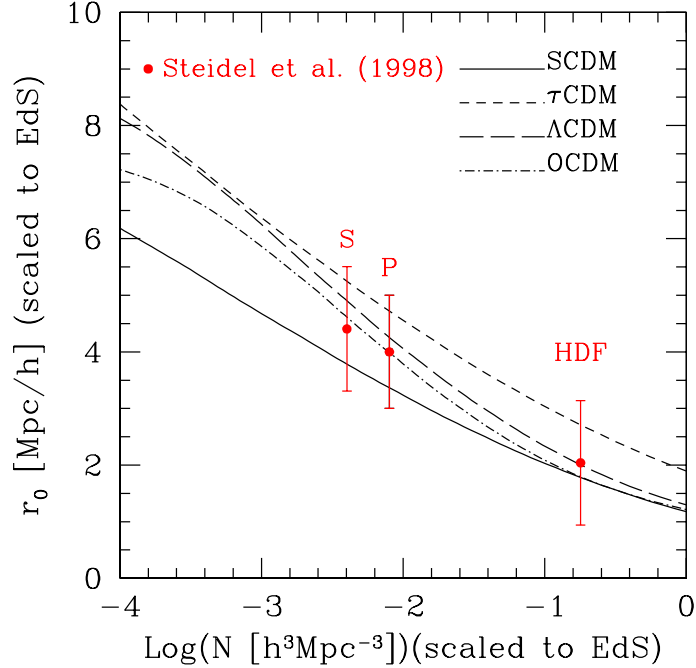}}
\caption{The scaled correlation length
[defined as the physical correlation length multiplied by 
$D_\star (1, 0)/D_\star (\Omega_0, \Lambda)$ where $D_\star(\Omega_0,
\Lambda)$ is the angular size distance 
at $z=3$] as a function of the scaled abundance
[defined as the comoving number density multiplied by
$\delta V(\Omega_0, \Lambda)/\delta V(1, 0)$,
where $\delta V((\Omega_0, \Lambda)$ is the comoving volume
per unit redshift at $z=3$].
Results are shown for four cosmogonies, as indicated
in the panel. Observational data for LBGs to three different
limiting magnitudes are taken directly from Steidel et al. (1998b).
}\end{figure}

The results presented above assume an LBG abundance equal to that
in the sample of Adelberger et al. (1998). Observations 
to fainter magnitude limits will give rise to denser samples, 
corresponding to less massive and less
biased halos. The correlation length is thus predicted to depend
on the sample selection criteria. We illustrate this in
Figure 3 where we plot the scaled correlation length as a 
function of the scaled abundance. (The directly observable abundance is
the number of LBGs per steradian and per unit redshift, so we normalize
the abundances predicted in each cosmology by the appropriate volume 
per unit redshift. In such a plot the position of the observed data is
independent of the cosmology assumed when analysing them.) We show 
theoretical predictions for four cosmologies with parameters similar
to those of the simulation set in Jenkins et al. (1998):
\begin{itemize}
\begin{enumerate}
\item SCDM:        $\Onow=1.0, \Lnow=0.0, h=0.5, \Gamma=0.5, \sigma_8=0.6$;
\item $\tau$CDM:    $\Onow=1.0, \Lnow=0.0, h=0.5, \Gamma=0.2, \sigma_8=0.6$;
\item $\Lambda$CDM: $\Onow=0.3, \Lnow=0.7, h=0.7, \Gamma=0.2, \sigma_8=1.0$;
\item OCDM:        $\Onow=0.3, \Lnow=0.0, h=0.7, \Gamma=0.2, \sigma_8=1.0$.
\end{enumerate}
\end{itemize}
The decrease in correlation length with increasing abundance is quite
strong and is very similar in the four cosmologies once they
are compared using the scaled variables. We show three observational
points taken directly from Steidel et al. (1998b).
The point at lowest abundance corresponds to the Adelberger et al.
(1998) data discussed above. The next point corresponds to a
reanalysis of the projected correlation data in Giavalisco et
al (1998). The point at highest abundance comes from an analysis
of LBG clustering in the Hubble Deep Field based on
photometric redshifts. Clearly all three points
are in good agreement with all the models. Again the data provide
little discrimination, but seem to provide strong
support both to the the hierarchical clustering picture and to the
identification of LBGs as the central galaxies of the most massive
haloes at $z\sim 3$. For example, if LBGs are assumed to be
objects undergoing short-lived bursts with a duty cycle of 10\%, then
the abundance of host haloes has to be 10 times the observed
LBG abundance. Figure 3 shows that the predicted correlation length 
would then be well below the value measured by Adelberger et al.
(1998). As these authors noted in their own analysis, improved 
measurements of LBG clustering should therefore place interesting 
constraints on the physical nature of these objects.

\subsection {A Model for the Structure of Lyman Break Galaxies}

 The simple model discussed in the last subsection relates the
abundance of LBGs to the mass of their haloes and so is able to
predict how they cluster. To study whether other properties of
the LBG population, for example their sizes, velocity dispersions and
star formation rates, are consistent with the inferred halo masses, we
need more detailed models for LBG formation. We develop such a model
here based on the disk formation model of Mo, Mao \& White (1998,
hereafter MMW)
and the phenomenological star formation laws of
Kennicutt (1998). In \S 3 we will apply this model to two specific
cosmologies, the $\tau$CDM and $\Lambda$CDM models discussed above.
The results of the last section can be used to scale the results
found there to other cosmologies of interest.

The disk formation model of MMW is an update of the scheme proposed
by Fall \& Efstathiou (1980, see also Dalcanton, Spergel \& Summers 1997, 
and references therein). This model reproduces the
properties of local disk galaxies well, and is consistent with the
observed evolution of disk galaxies out to redshifts  $\sim 1$ 
(Mao, Mo \& White 1998). The reader is referred to MMW
for details; here we only repeat the essentials of the model.
Briefly, after the initial protogalactic collapse the gas and dark
matter are assumed to be uniformly mixed in a virialized object with
density profile, 
\beq\label{profile}
\rho (r)={V_{\rm h}^2\over 4\pi G r^2}
{1\over \left[\ln(1+c)-c/(1+c)\right]}
{r/r_{\rm h}\over (r/r_{\rm h}+1/c)^2},
\eeq
where $V_{\rm h}$ and $r_{\rm h}$ are related to the
halo mass $M_{\rm h}$ by equation (\ref{VhM})
(Navarro, Frenk \& White 1996, 1997, hereafter NFW).
The quantity $c$ in equation (\ref {profile}) is known as the 
halo concentration factor and can be estimated for a halo
of given mass in any given cosmogony (see NFW).

As a result of dissipative and radiative processes,
the gas component gradually settles into a disk. We assume that
the mass of this disk is a constant fraction $\md$  
of the halo mass, and that its specific angular momentum is equal
to that of the dark halo. These assumptions are questionable but
they do reproduce the distribution of disk sizes and masses observed
at $z=0$ (see MMW). If the mass profile of the disk is taken to
be exponential, 
\beq \label{exp}
\Sigma(r) = \Sigma_0 \exp(-r/\Rd),
\eeq
then $\Sigma_0$, $\Rd$ and the galaxy's rotation curve
are determined uniquely. Specifically,
\bey\label{rd_sis}
\Rd &=& {1\over \sqrt{2}}\lambda r_{\rm h} F_R \nonumber \\
&\approx & 8.8 h^{-1}\kpc \left({\lambda \over 0.05}\right)
\left({V_{\rm h} \over 250\kms}\right)
\left[{H(z)\over H_0}\right]^{-1} F_R,
\eey
\bey\label{sig_sis}
\Sigma_0 &=& {M_d \over 2 \pi \Rd^2} \nonumber \\
&\approx & 380 {M_\odot \over {\rm pc}^2}
h\left({\md \over 0.05}\right) 
\left({\lambda\over 0.05}\right)^{-2}
\left({V_{\rm h} \over 250\kms}\right)
\left[{H(z)\over H_0}\right] F_R^{-2},
\eey
where $\lambda$ is the spin parameter of the halo and $F_R$ is a
constant of order unity (see Mao \& Mo 1998). From equations
(\ref{rd_sis}) and (\ref{sig_sis}) it is clear that 
high-redshift disks are generically smaller and denser than nearby
systems. For
example, for fixed $V_{\rm h}$ and $\lambda$, scalelengths are
8 times smaller at $z=3$ than at $z=0$ in an 
EdS universe; this factor is
about 4 for a flat universe with $\Onow=0.3$. 

In systems with $\lambda$ smaller than some 
critical value, $\lambda_{\rm crit}$, gas cannot settle to a
centrifugally supported disk without first becoming
self-gravitating. In this case collapse may be arrested by star formation
without formation of an equilibrium disk.
The size of the galaxy would then reflect the scale at which it became
baryon-dominated rather than its angular momentum support.
In such cases we assume the final size and density profile of the galaxy
to be those that it would have according to our disk model if its
halo spin were $\lambda_{\rm crit}$ rather than the actual value $\lambda$.
The final configuration is probably spheroidal rather than disk-like
in such systems, but this should not seriously affect the size and
velocity dispersion estimates given below. The effects on the star
formation rate are harder to assess, although we note that the 
phenomenological model we adopt does seem to describe nearby 
starbursts where conditions may be similar to those we envisage 
during the collapse of low spin systems (Kennicutt 1998). For detailed
modelling we will take $\lambda_{\rm crit}=m_d$, as discussed in MMW.

These assumptions, together with the star formation law which we discuss
next, allow us to compute the properties of a newly formed galaxy in any 
cosmogony for any given set of the parameters,
$M_{\rm h}$, $\lambda$ and $\md$.
As we will see below, most of our predictions do not depend on the 
exact value assumed for $m_d$, but some, like the star formation 
rate, vary strongly with this quantity. 

We estimate star formation rates using
the empirically based Schmidt law proposed by Kennicutt (1998):
\beq \label{sfr}
\Sigma_{\rm SFR} = 2.5\times 10^{-4} \left({\Sigma(r) \over 1 M_\odot {\rm
pc}^{-2}} \right)^{1.4} M_\odot {\rm~ yr^{-1}~ kpc^{-2}},
\eeq
where $\Sigma_{\rm SFR}$ is the SFR per unit area, and
$\Sigma(r)$ is the mass surface density of HI and ${\rm H_2}$
gas. This law fits the SFR in nearby galaxies over a total range of
more than $10^7$ in $\Sigma_{\rm SFR}$ with a scatter of a
factor of 2 or 3. We assume $\Sigma (r)$ to be given by equation (\ref{exp})
so our predicted SFR will scale approximately as $m_d^{1.5}$. 
In most of our calculations, we will take $m_d=0.03$, but this choice is 
quite arbitrary. In principle $m_d$ could range between 
0 and $f_{\rm B}\equiv\Omega_{\rm B}/\Omega_0$. A comparison
of nucleosynthesis calculations with the abundance of light elements
suggests $\Omega_{\rm B}\sim 0.02 h^{-2}$ 
(Schramm \& Turner 1998).
We assume $m_d\ < f_{\rm B}$, because condensation of halo gas into
LBGs is likely to be less than 100\% efficient. The uncertainty in the
appropriate value of $\md$ makes the SFR distribution the least robust
of our predictions. We will come back to this issue later. 

In hierarchical structure formation, haloes grow continuously by
accretion and merging. It is therefore important to examine
whether the gas associated with a dark halo has time to cool and 
settle into a disk before the halo merges into a larger system.
This question is best addressed
through numerical simulations. The formation of individual galaxies
in CDM cosmologies has been simulated by a number of authors (Navarro 
\& White 1994; Navarro, Frenk \& White 1995; Haehnelt, Steinmetz \& 
Rauch 1998; Navarro \& Steinmetz 1997; Weil, Eke \& Efstathiou
1998). These studies all agree that a large fraction of the gas
associated with high redshift halos is able to cool and condense into
disk-like systems. Furthermore the structure of these disks is
reasonably approximated by an exponential law over the bulk of the mass.
The outer disk is, however, frequently disturbed by
ongoing interactions and mergers, and this results in a cross-section
for damped Ly$\alpha$ absorption at $z\sim 2$ which is substantially
larger than predicted by an equilibrium disk model (Haehnelt, Steinmetz \&
Rauch 1998). A further problem is that these simulations all show
a substantial transfer of angular momentum from the cool gas to the
dark matter. This produces $z=0$ disks which are smaller than
observed, and so also smaller than our models predict. The resolution 
of this difficulty remains unclear (cf. Navarro \& Steinmetz 1997; 
Weil, Eke \& Efstathiou 1998). We persevere with our simple 
models because they {\it do} fit observations of nearby galaxies and they are
consistent with the observed evolution of disks out to $z\sim 1$. 

\section {Predictions for the LBG Population}

The simple galaxy formation model described above allows us to 
calculate the properties of the central galaxy in a halo with given 
mass, spin parameter and redshift.
To predict the properties of the LBG population, we also need to know
the distributions of $M_{\rm h}$ and $\lambda$ as a function of redshift. 
As in Section 2, we use the PS formalism to calculate
the abundance of dark haloes. $N$-body simulations show the distribution
of $\ln \lambda$ for dark haloes is approximately normal 
with mean ${\overline{\ln \lambda}}=\ln 0.05$ and dispersion
$\sigma_{\ln \lambda}=0.5$ (see equation [15] in MMW; 
Warren et al. 1992; Cole \& Lacey 1996; Lemson \& Kauffmann
1998). This distribution is found to depend only weakly 
on cosmology and on the mass and environment of haloes (Lemson 
\& Kauffmann 1998). With the distributions of $M_{\rm h}$ 
and $\lambda$ given, we can generate Monte Carlo samples 
of the halo distribution in the $M_{\rm h}-\lambda$ plane
at any given redshift. We can then use the galaxy formation model of
\S 2.2 to transform the halo population into an LBG population.

In order to model the observed LBG population, we must incorporate
the criteria by which they were selected. We will assume that the
colour and magnitude limits of the Steidel et al. sample result in
a set of $z\sim 3$ galaxies complete above some limiting star
formation rate. This neglects the fact that the magnitude limit
(${\cal R}<25.5$, corresponding to a $1500 {\rm \AA}$ luminosity limit, 
$L_{1500}\ga 1.3\times 10^{41} h^{-2} {\rm erg\, s^{-1} \AA ^{-1}}$
at $z=3$ in an Einstein-de Sitter universe)
may correspond to different SFRs in objects with differing dust
distributions or star formation histories (see \S 3.5). Since the
mean conversion from $L_{1500}$ to SFR is controversial,
we use the observed abundance of LBGs to set the limiting
SFR, checking {\it a posteriori} whether the resulting
value seems plausible. Specifically, we identify LBGs as the 
most rapidly star-forming galaxies at $z=3$, subject to the condition
that the comoving number density of the model 
population is the same as that observed. 
This defines a SFR threshold.
This procedure depends on the relation between SFR and
the far-UV luminosity only in a loose sense: it is valid
provided the uncertainties in the conversion from
SFR to far-UV luminosity do not induce a severe mixing 
between galaxies with different SFR. It also
requires that star formation in each LBG last 
for a period comparable to the Hubble time at $z\sim 3$;
otherwise the observed number of LBGs 
would be smaller than the number of haloes able 
to host them. As we will see in \S 3.5, this requirement 
is indeed fulfilled in our model.

Once the LBG population has been identified in this way in a Monte
Carlo simulation of the high-redshift galaxy population,
we can study its statistical properties in some detail.
For brevity, results are presented below only for the 
$\tau$CDM and $\Lambda$CDM models, and only for the abundance 
estimated by Adelberger et al. (1998). The results can be scaled
approximately to other models and other abundances by careful use 
of the results shown in \S 2.1.

\subsection {Halo circular velocities}
\begin{figure}
\epsfysize=9.5cm
\centerline{\epsfbox{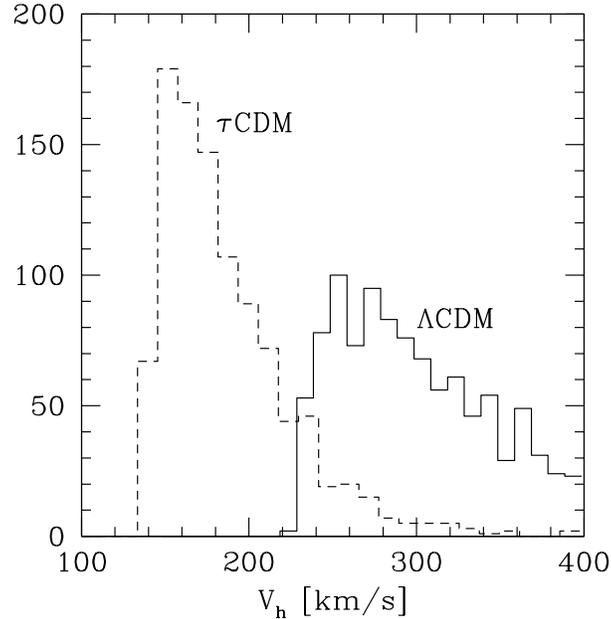}}
\caption{The circular velocity distribution for the
haloes which host LBGs. Solid and dashed histograms are 
for $\Lambda$CDM and $\tau$CDM, respectively.
}\end{figure}
Figure 4 shows the distribution of circular velocity 
for the haloes which host the LBGs.
In $\Lambda$CDM, these
circular velocities are quite big, with
a median of about $290\kms$, corresponding to 
a total halo mass $M_{\rm h}\sim 1.0\times 10^{12}
h^{-1}\msun$. In this model, LBGs are indeed associated 
with massive dark haloes. In contrast, the halo circular velocities in the 
$\tau$CDM model are much smaller. The median is now
about $180\kms$, corresponding to $M_{\rm h}\sim 
1.5\times 10^{11} h^{-1}$. In this cosmogony, relatively few massive
haloes form before $z=3$, and one has to include lower mass systems
in order to match the observed number density of LBGs.
Note that these median values are quite close to the lower
limits inferred for the halo circular velocities using the
simpler model of \S 2.1. In the current model lower mass halos can
make it into the sample if they have small $\lambda$ values, since
they are then predicted to be compact and to have higher than average
SFRs for their mass.

\subsection {Correlation functions}

At $z=3$ the characteristic mass of dark halos (defined by
$\Delta(M_*)=\delta_c(z)$) corresponds to a halo circular velocity 
$V_*\approx 160 \kms$ for $\Lambda$CDM and $V_*\approx 40 \kms$
for $\tau$CDM. Thus the haloes which host the LBG population
are much more massive than $M_*$, 
especially in $\tau$CDM, and the distribution of LBGs should be 
strongly biased relative to that of the mass. The predicted
bias factor can be obtained by averaging $b(V_{\rm h}, z)$, as given in
equation (\ref{bias}) over the $V_{\rm h}$-distribution
shown in Fig.4. The result is $b=2.8$ for $\Lambda$CDM 
and $b=5.0$ for $\tau$CDM. These are similar to the bias factors 
derived by Steidel et al. (1998a,b; see also
Adelberger et al. 1998) for the observed LBGs. 
\begin{figure}
\epsfysize=9.5cm
\centerline{\epsfbox{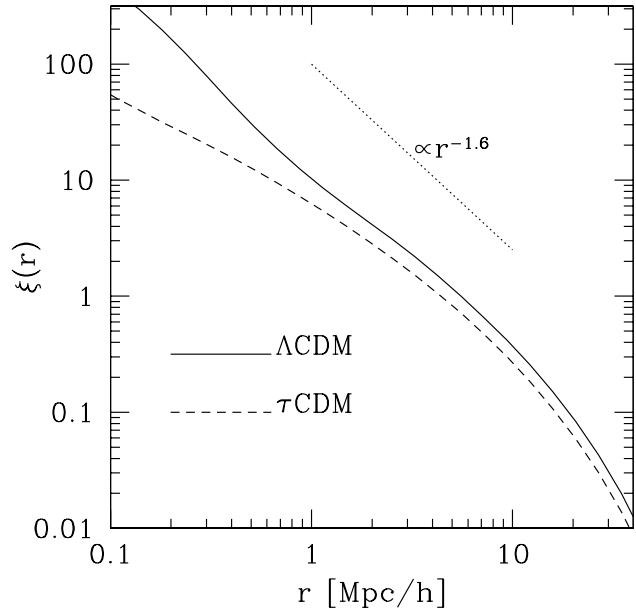}}
\caption{The predicted two-point correlation function for LBGs with the
abundance at $z= 3$ estimated by Adelberger et al. (1998). Results for
$\Lambda$CDM are shown by the solid line and $\tau$CDM  by the dashed line.
The dotted line is a power law $\xi\propto r^{-1.6}$.
}\end{figure}

Figure 5 shows the predicted correlation function for LBGs 
in the $\tau$CDM and $\Lambda$CDM models. In comoving units
the correlation length is $r_0\approx 4.5\mpch$ for $\tau$CDM, 
while $r_0\approx 5.5 \mpch$ for $\Lambda$CDM. 
These predictions are slightly below those of the simple model
of \S 2.1 as a result of the inclusion of some lower mass haloes.
They agree well with the observational result
of Adelberger et al. (1998) based on a count-in-cells analysis of
a fully spectroscopically confirmed sample.
This observational result may still be 
uncertain, however. For example, Giavalisco et al. (1998) used angular
correlation data for a larger and slightly fainter sample of LBGs
to infer $r_0=(2.1\pm 0.5)\mpch$ at $z=3$ in 
an Einstein-de Sitter universe. A reanalysis of these same data in
Steidel et al. (1998b) gave the intermediate abundance point plotted
in Figure 3 which is clearly in much better agreement both with the
models and with the Adelberger et al. point. Confirmation of
the measured amplitudes in independent data sets would clearly be
very valuable.

The present-day descendents of the LBGs will have a correlation
function which can be written as
\beq
\xi_0 (r)=b_0^2 \xi_{\rm m}(r,0),
\eeq
where $\xi_{\rm m}(r,0)$ is the present-day mass correlation function,
and the bias factor is related to that at the redshift $z$ where the
LBGs were identified by
\begin{eqnarray}
b_0(V_{\rm h}, z)
&=&1+{1\over \delta_c(z)}
\left[ {\delta_c^2(z)\over \Delta^2 (M_{\rm h})}-1\right]
\nonumber\\
&=&1+{g(z)\over (1+z)g(0)}\left[b(V_{\rm h},z)-1\right]
\end{eqnarray}
(Mo \& White 1996; Mo \& Fukugita 1996). Note that each present-day
descendent must be weighted by the number of LBGs it contains when
estimating these correlations. The detailed models of Baugh et al.
(1998) suggest that most descendents contain only a single LBG. 
Using the values of $b$ estimated above, we obtain
$b_0=1.6$ for $\Lambda$CDM and $b_0=2.0$ for $\tau$CDM.
Normal bright galaxies at $z=0$ have 
$b\sim \sigma_8^{-1}$. The correlation amplitude for LBG descendents
is thus predicted to exceed that of normal bright galaxies by a factor
of about $(\sigma_8 b_0)^2$, or $\sim 2.5$ for $\Lambda$CDM
and $\sim 1.4$ for $\tau$CDM. This is consistent with the
conclusions of Mo \& Fukugita (1996), Baugh et al. (1998) and Governato
et al. (1998); LBG descendents are primarily among the brightest
galaxies and are found preferentially in clusters.

\subsection {Half-light radii}
\begin{figure}
\epsfysize=9.5cm
\centerline{\epsfbox{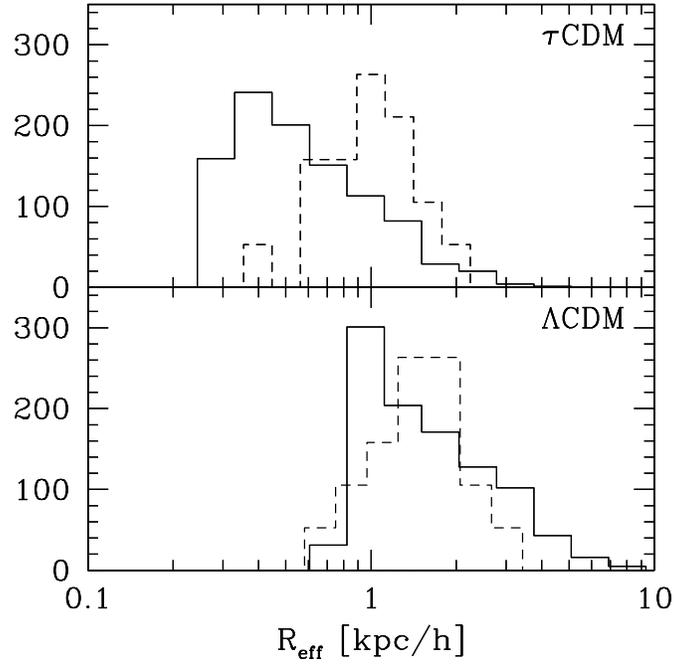}}
\caption{The distribution of half-light radii for LBGs. 
In each panel the solid histogram gives the model prediction and the
dashed histogram shows observational data (cf. table 2 in Giavalisco,
Steidel, \& Macchetto 1996 and table 2 in Lowenthal et al. 1997).
}\end{figure}
\begin{figure}
\epsfysize=9.5cm
\centerline{\epsfbox{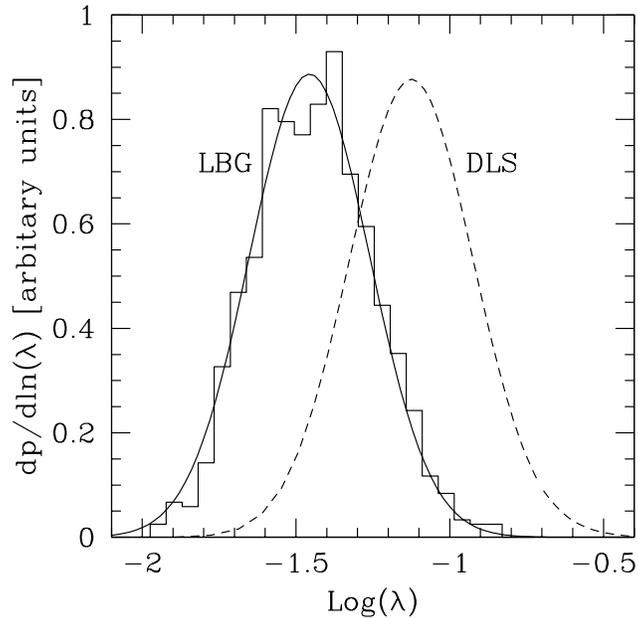}}
\caption{The spin parameter distribution for the haloes which
host LBGs (solid curve) is compared to that for haloes
which host damped Ly$\alpha$ absorption systems (DLSs, dashed curve).
Note that while LBGs are biased towards haloes with low spin,
DLSs are biased towards haloes with high spin.   
}\end{figure}
Figure 6 shows the predicted distribution of effective radius, $R_{\rm
eff}$, for the LBG population. We define $R_{\rm eff}$ as the
semimajor axis of the isophote which contains half of the star
formation activity. This is easily calculated from our model and 
will coincide with the size of the region containing half the
observed light provided the relation between SFR and far-UV surface 
brightness does not vary much across a galaxy. 
Half-light radii are predicted to be quite small. The median
$R_{\rm eff}$ is about $2\kpch$ for $\Lambda$CDM , while it is only
$0.6\kpch$ for $\tau$CDM. This large difference arises because
host halos are less massive and $H(z)/H_0$ is larger in
in $\tau$CDM than in $\Lambda$CDM [see equation (\ref{VhM})].
Although partially offset by the fact that angular size
distances are larger in the low density case, it opens the possibility
that size measurements for LBGs might significantly constrain 
cosmological parameters.

HST imaging of LBGs (to a magnitude limit comparable to ${\cal R}=25.5$)
in the Hubble Deep Field by 
Lowenthal et al. (1997) gave values of $R_{\rm eff}$ in the
range $(0.5\to 2.1)\kpch$, with a median near $1.1\kpch$, under the
assumption of an Einstein-de Sitter universe.
The corresponding range is $(0.8\to 3.6)\kpch$, with a median near 
$1.8\kpch$, for a flat universe with $\Lambda=0.3$. Similar
results were obtained by Giavalisco, Steidel \& Macchetto (1996).
These data agree well with our models for $\Lambda$CDM, but 
are perhaps somewhat larger than predicted for $\tau$CDM. 
The observational data are still quite sparse, and larger and more
complete samples are needed to get reliable constraints.

The small sizes predicted for LBGs may appear surprising given the
large circular velocities predicted for their halos.
For given $V_{\rm h}$, halo size decreases with $z$
as $H^{-1}(z)$ (cf. equation \ref{VhM}). At $z=3$ this gives a 
factor of 8 in an Einstein-de Sitter universe and a factor of $4$ in a flat 
universe with $\Omega_0=0.3$. In our model, the ratio of galaxy size
to halo size depends only on $\lambda$ and is independent of $z$.
The small sizes of the LBGs are due mostly to the small size of their
haloes, but also to the fact that, since we select LBGs according
to SFR, they are biased towards haloes with small 
$\lambda$; smaller $\lambda$ gives higher surface density (and so higher SFR)
but smaller size. This bias is most clearly shown
in Figure 7, where we show the spin parameter distribution for the LBG
population. The distribution of $\ln \lambda$ is approximately normal
with ${\overline{\ln \lambda}}=\ln 0.035$ and
$\sigma_{\ln \lambda}=0.4$. This should be compared with the 
original distribution which had ${\overline{\ln \lambda}}=\ln 0.05$ and
$\sigma_{\ln \lambda}=0.5$.

\subsection {Stellar velocity dispersions}

\begin{figure}
\epsfysize=9.5cm
\centerline{\epsfbox{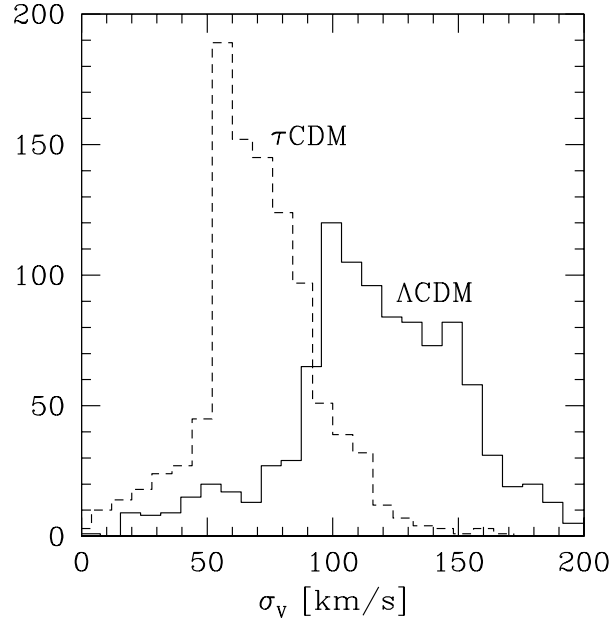}}
\caption{The line-of-sight stellar velocity dispersion distribution of LBGs. 
Solid and dashed histograms are 
for $\Lambda$CDM and $\tau$CDM, respectively.
}\end{figure}
Although our models predict the circular velocities of the haloes
hosting LBGs to be quite big (cf. Fig. 4), the stellar velocity 
dispersions of the LBGs themselves can be substantially smaller.
This is a result of a combination of projection effects with the fact
that the observed stellar distribution samples only the very central
regions of the halo potential well. To show this we construct
the distribution of SFR-weighted line-of-sight velocity dispersion 
for our Monte Carlo samples of model LBGs. For a disc galaxy seen at
inclination $i$, this dispersion is defined as 
\beq
\sigma_V^2={1\over 2 M_{\rm SFR}}
\int _0^{\infty} V_{\rm rot}^2(r)\sin^2(i)~
\Sigma_{\rm SFR}(r) 2\pi r dr,
\eeq
where the star formation surface density $\Sigma_{\rm SFR} (r)$ is 
obtained from equations (6) to (9), 
$M_{\rm SFR} \equiv \int_0^\infty \Sigma_{\rm SFR} (r) 2\pi rdr$, 
$V_{\rm rot}(r)$ is the disk rotation curve,
and $i=0$ refers to a face-on disc. In our Monte Carlo samples we
assume $\sin (i)$ to be uniformly distributed on $[0,1]$ corresponding
to randomly oriented galaxies. For the `spheroid' population
discussed in Section 2.2 (i.e. for systems with 
$\lambda<\lambda_{\rm crit}$), we assume the stars to be in random 
motion, with line-of-sight velocity dispersion given by 
\beq
\sigma_V^2={1\over 3M_{\rm SFR}}
\int _0^{\infty} V_{\rm rot}^2(r) \Sigma_{\rm SFR}(r) 2\pi r dr,
\eeq
where $V_{\rm rot} (r)$ and $\Sigma_{\rm SFR} (r)$ are calculated as for disks
but assuming $\lambda=\lambda_{\rm crit}$ rather than the true value.
Figure 8 shows the resulting $\sigma_V$ distribution for our Monte
Carlo
samples of LBGs. Comparing this figure with 
Fig.4 it is clear that stellar velocity dispersions 
are typically much smaller than halo circular velocities. 
The median values of $\sigma_V$ are $\sim 120\kms$
for $\Lambda$CDM and $\sim 70\kms$ for $\tau$CDM.
Much of this reduction is a result of the bulk of star formation
occurring on the inner rising part of the disk rotation curve.
The values would thus be even smaller if observations were
assumed to sample only the inner part of the light distribution.
These results suggest that even if stellar velocity dispersions could
be reliably measured for LBGs, it would be difficult to use them to 
infer the mass of the associated halo.

Based on the Ly$\alpha$ emission line widths observed in
six LBGs, Lowenthal et al. (1997) inferred  
$100\kms <\sigma_V <230 \kms$ with a median $\sim 140 \kms$. On the
basis of this they argued that LBGs are likely to be
the low-mass, star-bursting building blocks of present-day 
galaxies. A rather different interpretation appears natural in
our model. The observed velocity dispersions agree well with
what we would predict for LBGs in a $\Lambda$CDM cosmology, and 
appear too large to be consistent with $\tau$CDM.
These values would seem to require LBGs to be the central galaxies
in haloes with mass $\sim 10^{12}\msun$. It would be wrong to put much
weight on this conclusion, however, since it is now clear that the
widths of the Ly$\alpha$ lines in LBGs are often substantially
affected by radiative transfer effects and by non-gravitational motions
in the emitting gas; as a result they may bear little relation to the
stellar velocity dispersion of the underlying galaxy (Pettini et al.
1998).

Measurement of emission line widths in the near infrared may provide
a more reliable estimate of the virial velocity dispersion within
LBGs. Results for five galaxies are reported by 
Pettini et al. (1998) based on UKIRT observations of the
${\rm H}\beta$ and [OIII] emission lines. For four galaxies they
find $\sigma_V\sim 70 \kms$, while for the fifth $\sigma_V\sim 200
\kms$. The lower values agree well with our $\tau$CDM predictions, 
but seem on the small side to be consistent with $\Lambda$CDM. 
On the other hand, a value as large as $200 \kms$ is predicted to be
very rare in $\tau$CDM and also quite unusual in $\Lambda$CDM.
Clearly the observed sample is too small to draw reliable
conclusions, and the relation of these linewidths to the underlying
stellar velocity dispersion, while probably simpler than that for the
Ly$\alpha$ line, is still open to question. In this context, it is
interesting to note that in nearby starburst galaxies the velocity
dispersions inferred from forbidden emission lines
are substantially smaller than the rotation velocities
of the host galaxies (e.g. Lehnert \& Heckman 1996).
This appears to reflect the concentration of star formation to the 
nuclear regions where the rotation curve is still rising, and so
is an example of the effect which leads us to predict LBG velocity
dispersions much smaller than the circular velocities of their haloes.

\subsection {SFR functions}
\begin{figure}
\epsfysize=9.5cm
\centerline{\epsfbox{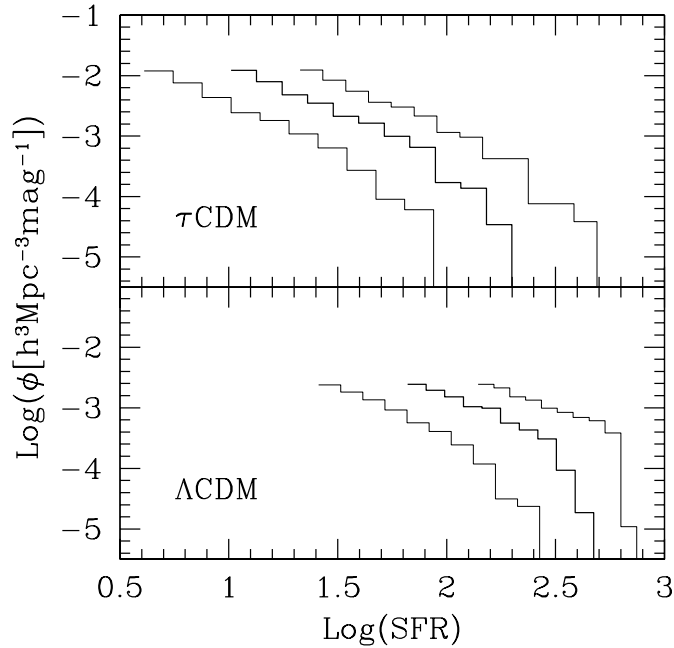}}
\caption{The star formation rate function (proportional to the UV luminosity
function) for LBGs in our $\Lambda$CDM and $\tau$CDM models. The bold
histogram in each case is for our standard disk mass fraction
$m_d=0.03$. The two light histograms are
for $\md=0.015$ (left) and $\md=0.06$ (right), respectively.
}
\end{figure}
The UV spectra of LBGs are dominated by the integrated continuum of
O and early B stars. Since these stars have short lifetimes,
the observed UV luminosity is directly determined by the star
formation rate. The main uncertainties in the conversion between these
two quantities come from the poorly constrained shape of the stellar
Initial Mass Function and, especially, from the difficulties in
establishing appropriate corrections for the amount of obscuration by
dust. For given halo mass, halo spin parameter and disk mass fraction, 
our model uniquely predicts the disk surface density (see
equations [\ref{exp}-\ref{sig_sis}]). The star formation rate can then 
be determined as a function of radius (and so integrated over the
galaxy as a whole) using equation (\ref{sfr}) under the assumption
that the disk is (at least initially) fully gaseous. Thus we can
determine the abundance of galaxies according to their SFR in our
Monte Carlo samples of LBGs. Such functions are shown in Figure 9 
for $\Lambda$CDM and $\tau$CDM, and for three choices of the disk mass
fraction. 

Figure 9 shows clearly that the SFR functions depend strongly on $m_d$. This
is in contrast to the properties discussed in earlier sections which
vary quite weakly with $m_d$ over the range considered here --
with increasing $m_d$ galaxies become somewhat more compact and their
internal velocities increase. If we take $\md=0.03$, the limiting
SFR for $\tau$CDM is about 10$\my$ and values up to 200$\my$ are
found in our Monte Carlo samples. Doubling $\md$ increases the cut-off
SFR to 30$\my$ and the largest values to $\sim 400\my$. Larger values
are found in the $\Lambda$CDM case. For $\md=0.03$ the cut-off occurs
near 60$\my$ and the largest values approach 500 $\my$. For $\md=0.06$
the corresponding numbers are 120$\my$ and 800$\my$.
The larger rates for $\Lambda$CDM reflect the fact that LBGs are 
hosted by more massive haloes in this model. 

For comparison, for a standard IMF and without {\it any} correction
for obscuration the UV luminosities of the LBGs studied by 
Steidel et al. (1998a,b) correspond to star formation rates in the range
$4-50\my$. The corrected rates depend significantly on the extinction
law assumed and on the detailed interpretation of the data. Steidel et
al (1998) obtained a mean correction of 2.0 and a range of corrected
values $4-120\my$ for an assumed SMC extinction law, while for the
extinction law of Calzetti (1997) they found a mean correction of 7.7 
and a range $4-1500\my$. Pettini et al. (1998) concluded from their IR
observations of a few galaxies that the second of these corrections
may be closer to the truth (see also Meurer 1997). All the rates
quoted here assume an Einstein-de Sitter cosmology with $h=0.5$. For 
our $\Lambda$CDM parameters they need to be increased by a factor of 2.5.
Figure 9 shows that any of these ranges can be accommodated in either
of our models for a plausible choice of $\md$. Recall that taking
$\md$ to be constant is a simplification of convenience,
and that in practice a range of values (including, perhaps, a
systematic dependence on halo mass) is to be expected. Once $\md$ is
chosen to match the observed limiting UV luminosities, the 
SFR functions in Fig. 9 give a good fit to the LBG luminosity function
constructed by Dickinson (1998).

If we divide the disk masses of our model LBGs by their inferred SFRs
the resulting decay time constant for the star formation is typically
30\% of the age of the universe at $z=3$. This is comparable to the
timescale on which the LBG halos double their mass, and so to that
on which new gas can be supplied. Thus there does not appear to be
a ``gas supply'' problem in the models, and there is no need to
invoke a bursting mode for the observed star formation 
(cf. Lowenthal et al. 1997; Somerville, Primack \& Faber 1998). 

\subsection{Connection to high-redshift damped Ly$\alpha$ systems}

As can be seen from Fig 7, the LBG population in our model is biased 
towards objects with small angular momentum; these have higher surface
densities and so higher star formation rates. In contrast the
cross-section for damped Ly$\alpha$ absorption by equilibrium disks
is dominated by objects with large angular momentum, since the
cross-sections of individual objects scale as 
$\pi \Rd^2 \propto \lambda^2$ (see MMW). We illustrate this 
in Fig.7 by overplotting the distribution of $\lambda$ predicted for a
population of equilibrium disks selected according to their damped
Ly$\alpha$ cross-section. As emphasised most recently by Prochaska
\& Wolfe (1997), matching the total observed cross-section for 
damped absorption in this model seems to require the inclusion of 
systems with rotation speeds too small to be consistent with the
observed velocity widths of the associated low ionization metal line systems.
Haehnelt, Steinmetz \& Rauch (1998) demonstrate that this discrepancy
is plausibly eliminated when proper account is taken of the fact that
hierarchical formation models predict the outer
parts of many high-redshift disks to be both spatially and kinematically 
distorted by interactions and  mergers. This increases both their
cross-sections and their line widths. Since the susceptibility to
tidal distortion increases with disk size (e.g. Springel \& White
1998) it is reasonable to suppose that the bias of such tidally
distorted systems towards large $\lambda$ is at least as strong as our
predictions for equilibrium disks

The median spin parameters for the LBGs and the damped Ly$\alpha$
systems (DLSs) in Fig 7 are 0.035 and 0.08,
respectively. There are essentially no LBGs with $\lambda>0.1$ while
nearly $\sim 40\%$ of DLSs have $\lambda>0.1$. 
Even without accounting for the loss of gas through star formation,
the total cross-section of the LBG population for damped absorption
is only about one fifth of that of the DLS population to the same mass
limit. Since the star formation rate 
per unit area is $\propto \Sigma^{1.4} 
\propto \lambda^{-2.8}$, the SFR per unit area in DLSs is, 
on average, a factor of $\sim 10$ lower than in LBGs. 
As a result most DLSs will not resemble the
bright and compact LBGs, although some may be detected as gas discs
with faint LBGs at their centres. We thus predict LBGs and DLSs to
be quite distinct populations. 
Because of their more rapid star formation,
LBGs should have systematically higher metallicity and dust
content than DLSs. Simple attempts to make direct connections 
between these two populations are therefore dangerous.
For example, it may be misleading to compare
the metallicities of DLSs directly with those of LBGs
(Madau, Pozzetti \& Dickinson 1998).
   
\subsection {Discs or spheroids?}

In Section 2.2 we suggested that LBGs in high spin haloes
may be rotationally supported disks,
while those in low spin haloes
may be (partially) supported by random motions. 
Stars in the latter systems may form before the gas can settle
to centrifugal equilibrium and so may produce spheroids. 
In reality, we  might expect to see the
whole spectrum from completely rotationally supported
discs, through partially rotationally supported disc/bulge systems, to 
random-motion supported spheroids. This would be reflected 
in a variety of shapes in the images
of LBGs. If it is correct, as we have argued, that observed LBGs are
predominantly in low-spin haloes, they should be biased
towards spheroids, and so should appear more compact and
less flattened than the general population. According to our crude
stability criterion, the fraction of the 
population in ``spheroids'' is given roughly by the condition 
$\lambda<m_d$. Thus, if $m_d>0.035$, the majority
of LBGs may be spheroidal. Unfortunately, these arguments are quite
sketchy, and reliable quantitative predictions will require detailed and
convincing simulations of how gas settles and forms stars in these
objects.

\section {Discussion}

In this paper, we have modeled both the clustering and the internal 
structure of the recently discovered population of Lyman
break galaxies. The assumption that these objects
are the central galaxies of massive halos allows models which fit the
structural and star formation properties of nearby spirals to be
scaled to $z\sim 3$. The populations observed by
Steidel et al. (1998a,b) and Adelberger et al.
(1998) can then be identified as the most rapidly star-forming (and 
hence brightest) galaxies, and the conversion between star formation 
rate and UV luminosity can be set so that the predicted LBG abundance 
equals that observed. For given cosmological parameters
the models then predict distributions of size, velocity
dispersion, and star-formation rate for the observed samples, as well as
the strength of their clustering. A reasonable fit to the current
data can be found in all currently popular cosmologies.

The models predict that Lyman break galaxies and damped Ly$\alpha$
absorbers should form almost disjoint populations. These two
kinds of object are currently the best available tracers of
galaxy formation at high redshifts. However, LBG samples are biased
in favour of systems of low angular momentum, and so high star formation
rate, while DLS samples are biased in favour of systems of high
angular momentum, and so large absorption cross-section. This results
in substantial systematic differences in size, kinematics and star 
formation history. Most LBGs should be compact, high surface
brightness, possibly spheroidal systems, whereas most DLSs should be 
extended, low surface brightness, rotationally supported disks. The
difference in star formation rates will plausibly cause the LBGs to 
have significantly higher metal abundances than the DLSs. Direct
observation of these differences would
provide important evidence for the picture we have developed.

Because we do not attempt any explicit modelling of the cooling 
and feedback processes associated with galaxy formation, we treat
the mass fraction which settles to the halo centre 
as an adjustable constant. In addition we assume the specific angular 
momentum of the central galaxy to be the same as that of its halo.
These are probably the most questionable aspects of our models. In
Mo, Mao \& White (1998) we showed that these simple assumptions
provide a surprisingly good fit to the properties of local
spirals, and in Mao, Mo \& White (1998) that they are also consistent
with the available data on the evolution of disks out to $z\sim 1$. 
This does not, of course, imply that they must be good assumptions
also for LBGs at $z\sim 3$. Our predictions for sizes, dispersion
velocities and clustering depend only weakly on 
disk mass fraction, and the clustering predictions are also nearly
independent of the assumed specific angular momentum. On the
other hand, the predicted star formation rates depend strongly on
both assumptions. The observed clustering of LBGs should thus be
considered a robust confirmation of hierarchical galaxy formation
theory and of the assumption that the LBGs in current spectroscopic 
samples are primarily the dominant galaxies of the most massive 
haloes at $z\sim 3$. Our predictions for sizes and velocity dispersions
are also reasonably robust provided angular momentum transfer is no
more efficient in LBG formation than at low redshift. The agreement 
between models and observational data provides significant further
confirmation of the basic paradigm. For the star formation rates we
are reduced to noting that the observed UV luminosity functions can be
matched for any of the mean obscuration factors currently under debate
using physically plausible values of $\md$, the LBG mass fraction.

The model we propose differs significantly from the low-mass
starburst scenario suggested by 
Lowenthal et al. (1997). In this scenario,
star formation in each LBG lasts for less than $10^{8}$ yr,
so the observed objects are only a small 
fraction ($\la 10\%$) of the total (bursting plus dormant) 
population. If the abundance of potential LBGs is
at least ten times the value we have adopted, and if again 
LBGs are assumed to be the dominant objects in their haloes, then
substantially lower mass haloes must be able to host LBGs. This results
in reductions of the limiting halo circular velocity (and so of the
predicted sizes and velocity dispersions) by a factor of 
2 or more, and of the predicted clustering length by a factor of 
at least 1.6. Such reductions make the apparent agreement with
observation significantly worse, particularly for $\tau$CDM where
they appear to be excluded by the observed sizes and velocity
dispersions. Clearly better determinations of $r_0$, $R_{\rm eff}$ 
and $\sigma_V$ should decide definitively between the two models.

In recent work Somerville, Primack \& Faber (1998) have suggested 
a variant of this starburst picture in which observed LBGs
are no longer assumed to be the dominant galaxies in their haloes
but are taken to be satellite systems 
undergoing interaction-induced starbursts 
in relatively massive haloes.
Our modelling suggests several
difficulties with this picture. We would again predict the sizes and 
velocity dispersions of these low-mass objects to be quite small in 
comparison with the current data. In addition, if on average there is
one bursting object per massive halo (so that large-scale clustering
is the same as in our model and so in agreement with the
data) then Poisson statistics predict that more than half the LBGs
should share their halo with a second observable object, whereas the
number of close pairs ($\Delta r_p \la 100\kpch$) is very small
in the observed samples (Giavalisco et al. 1998). If bursting satellites
can occur in lower mass haloes, thus reducing the mean number per halo
and so the probability of multiple objects, then the clustering strength
is also reduced as before. Finally, current semi-analytic models
suggest that in galaxy mass halos the amount of fuel available for star
formation in the central object is considerable larger than that
available in all the satellites combined (Kauffmann et al. 1993; Baugh
et al. 1998; Governato et al. 1998). A bursting satellite model
then requires a substantial fraction of the observed star formation 
at $z\sim3$, and in particular the most luminous starbursts, to be
occurring in objects with a small fraction of the total available
fuel.

In conclusion, a model in which current spectroscopic samples of
Lyman break galaxies are dominated by the central galaxies of the most
massive haloes at  $z\sim 3$ seems to account in a simple and consistent
way for the sizes, velocity dispersions, star formation rates and
clustering of these objects. The current rather sparse data appear to
favour such models over alternatives in which the galaxies are assumed
to be undergoing short-lived starbursts.

\section*{Acknowledgments}

We thank Chenggang Shu, Chuck Steidel
and Art Wolfe for many useful discussions
on the topics discussed in this paper. 
This project is partly supported by
the ``Sonderforschungsbereich 375-95 f\"ur Astro-Teilchenphysik der
Deutschen Forschungsgemeinschaft''. 

\vfill\eject
{}

\end{document}